# Inversion of lattice models from the observations of microscopic degrees of freedom: parameter estimation with uncertainty quantification

Mani Valleti,[1,4] Lukas Vlcek,[2,3] Rama K. Vasudevan,[4] and Sergei V. Kalinin[4]

[1] Bredesen Center for Interdisciplinary Research, University of Tennessee, Knoxville, TN 37996, USA

[2] Materials Science and Technology Division, Oak Ridge National Laboratory,
Oak Ridge TN 37831, USA

[3] Joint Institute for Computational Sciences, University of Tennessee, Knoxville, Oak Ridge, TN 37831, USA

[4] The Center for Nanophase Materials Sciences, Oak Ridge National Laboratory, Oak Ridge, TN 37831, USA

**Abstract**

Experimental advances in condensed matter physics and material science have enabled ready access to atomic-resolution images, with resolution of modern tools often sufficient to extract minute details of symmetry-breaking distortions such as polarization, octahedra tilts, or other structure-coupled order parameters. The patterns of observed distortions in turn contain the information on microscopic driving forces defining the development of materials microstructure and associated thermodynamics. However, the analysis of underpinning physical models from experimentally observed microscopic degrees of freedom remains a largely unresolved issue. Here, we explore such an approach using the paradigmatic Ising model on a square lattice. We show that the microscopic parameters of the Ising model both for ferromagnetic and antiferromagnetic case can be extracted from the spin configurations for temperatures an order of magnitude higher than the phase transition and perform uncertainty analysis for such reconstructions. This suggests that microscopic observations of materials with sufficiently high precision can provide information on generative physics at temperatures well above corresponding phase transition, opening new horizons for scientific exploration via high-resolution imaging.

Progress in condensed matter physics and materials science over the last century is inherently linked with the development of simple models that can capture specific aspects of materials physics. A special place among these belongs to lattice models. In these, the material is represented as a system of interacting units (or spins) on a spatial lattice. Depending on the possible spin states and type of interaction, multiple classes of lattice models such as Ising, Heisenberg, etc. emerge. Once the lattice and character of interactions are defined, the macroscopic observables such as average magnetization, heat capacities, and energies can be determined. These can be further used to construct macroscopic phase diagrams delineating the regions of dissimilar ordering in the temperature-parameter spaces[1-3].

Traditionally, exploration of the lattice models was performed via numerical (and in rare case analytical) methods. Over the last several years, machine learning techniques were shown to be a powerful tool in analyzing the nature of the ground states and phase diagrams of the lattice models, giving rise in rapidly growing effort in this field.[4, 5] For example, neural networks and even more basic machine learning methods have been shown to have a high degree of success in classification of distinct phases within Ising-type models when trained directly on the snapshots of the configurations. A connection between a certain class of deep learning models and the renormalization group was also found and has a strong mathematical link.[6-9] However, these approaches essentially allow only to analyze the intrinsic behavior of the model. The more perplexing issue of correspondence between the model and a realistic solid remains open.

In fact, establishing the relationship between real materials and the parameters of the lattice model, including the type of lattice and interactions (class of model) and the strength of corresponding interactions remained a key issue since the early days of the statistical physics of solids. Traditionally, such a comparison was performed through the measurements of macroscopic thermodynamic properties suggesting the presence and character of phase transitions and variable temperature scattering studies, that, when combined with the knowledge of materials structure and collection of phenomenological rules, allowed to suggest the possible universality class. Correspondingly, the numerical model parameters are then derived via the comparison of experimentally observed parameters (e.g. temperatures of phase transitions) to the phase diagrams of corresponding models.

Several approaches for determining the microscopic model parameters via mesoscopic observables have been proposed. Ovchinnikov and Kumar[10, 11] demonstrated the use of the feed forward neural network trained on the theory data to extract parameters of the Ising model. Similarly, Li et al.[12] proposed the use of perturbing a system and measuring the dynamic response of mesoscopic proxy quantities (where order parameters cannot be measured directly) as an indicator of structural phase. Clustering on these dynamic responses leads directly to an interpretable phase diagram in the bias-temperature space and was shown via both experiment and simulation to be applicable. However, in these approaches the microscopic degrees of freedom are accessed solely via mesoscopic proxies, with implicit assumptions that microscopic degrees of freedom are thermalized, and parameters of the measurement system are known sufficiently well to avoid reconstruction errors. Similarly, the analysis of the uncertainties associated with such an analysis was not attempted.

Recently, the resolution and precision in imaging techniques, notably Scanning Transmission Electron Microscopy, has achieved the level sufficient for determining the position of atomic columns at the ~pm level.[13] This, in turn, allowed extraction of the order parameter fields such as polarization,[14-17] octahedra tilts,[18-20] or chemical expansion[21, 22] on a single unit cell level. These in turn have been connected to the mesoscopic physics via the analytical or numerical solution of the mesoscopic phase field models for simple interface or topological defect geometries, allowing to extract parameters such as gradient energy[23] and flexoelectric terms.[24] However, the wealth of information contained in observed microscopic degrees of freedom remained largely unexplored.

It has been proposed that energy parameters of lattice models can be obtained directly from the observations of the corresponding microscopic degrees of freedom via statistical distance minimization method.[25] This approach can be compared to analysis of time fluctuations in techniques such as single force spectroscopy,[26] where time-resolved fluctuations can be used to extract the free energy surface of a material along a certain transition path. Similarly, observation of composition fluctuations can be used to reconstruct the interaction parameters in a non-ideal solid solution, opening pathways to explore the generative model.[27, 28] However, the general applicability domain of such analyses, as well as associated reconstruction errors, remained unexplored. For example, questions such as the role of disorder in the inference, and the specific parameter ranges in which maximal information on the system can be extracted are of obvious importance for experimental design and data analysis.

Here, we explore the applicability of statistical distance reconstruction for the paradigmatic 2D Ising model for a square lattice over a broad parameter space. We demonstrate that this approach, provided the *a priori* knowledge of a model, allows to determine the thermodynamic parameters well above the corresponding transition temperatures. We further perform uncertainty quantification of this approach, establishing the model behavior for an idealized case of known generative model.

As a model, we have chosen the classical Ising Hamiltonian model realized on $N^2$-evenly spaced square lattice sites. Each lattice site is identified with a specific spin state ($\sigma$) that can adopt two values, $\sigma \in \{-1, 1\}$. The total energy for a square lattice in the absence of external field with a specific configuration is given by the Ising Hamiltonian function:

$$H(\sigma) = -\sum_{<i,j>} J_1\, \sigma_i \sigma_j \qquad (1)$$

Where, $H$ is the Hamiltonian of the given configuration, $J_1$ is the isotropic interaction parameter corresponding to the nearest neighbors, and <i, j> indicates the sum over all the nearest neighbors The model also incorporates bond-disorder in which the magnitude of interaction parameter at each lattice site is normally distributed around a given $J_1$ with a given standard deviation.

Monte Carlo simulations were performed on a domain of N * N lattice sites. At each step, spin configuration of each lattice site was attempted to be reversed: when energetically favorable the spin was reversed; when unfavorable the spin-flip acceptance criteria were determined by the Metropolis algorithm. In such cases, the probability of the spin flip is determined by Boltzmann distribution and is given by

$$P_\beta(\sigma) = e^{-\beta H(\sigma)} \qquad (2)$$

where $\beta$ is the inverse temperature ($1/k_B T$). The system was equilibrated for 1000 steps prior to actual data collection. Macroscopic observables including energy, specific heat, net magnetization and susceptibility were computed by using the individual properties of microstates on the subsequent 1000 steps. All the macroscopic properties are calculated per site to get intensive values.

The base case results for an interaction parameter $J_1 = 0.65$ with a bond-disorder of 10% shows a typical magnetization behavior (Figure 1a). The classical phase transition from ferromagnetic (net absolute magnetization ~ 1) to paramagnetic (net absolute magnetization ~ 0) is observed at a reduced temperature, $T_r \sim 1.6$. This phase transition is also evident from the susceptibility and specific heat curves (Figure 1b and d).

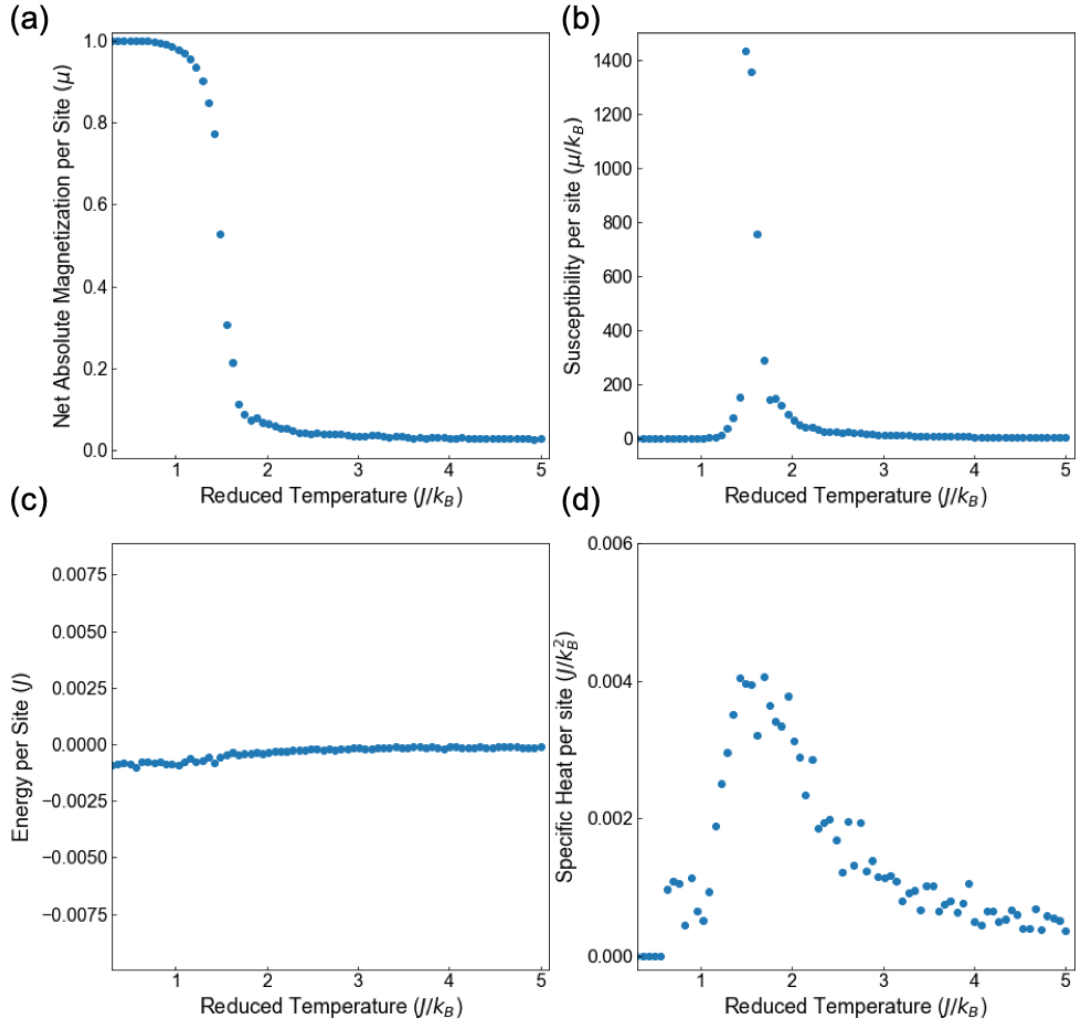

Figure 1. (a) Net absolute magnetization per site, (b) Susceptibility per site, (c) Energy per site and (d) Specific heat per site as a function of reduced temperature for the base case ($J1 = 0.65$, bond disorder = 0.1 and lattice size = 40*40)

Simulations as described by equation (1) yield the macroscopic parameters of the system. The general trends observed for these simulations are highlighted in contour plots, Figure 2(b-d). Transition behavior are observed for ferro-magnetic and anti-ferromagnetic domains in the specific heat contour. Net average magnetization and susceptibility do not show the transition for anti-ferromagnetic cases. The former occurs due to the checkerboard pattern of spin states that lead to essentially a zero net-magnetization. We note that comparison of the absolute values of these at any given temperature with the experiment is highly non-trivial, due to the presence of normalization factors, etc. However, at each temperature the system is also characterized by a certain pattern of spins, as shown in Figure 2(a, e). While these patterns are stochastic in nature and have no local return point memory (i.e. will not be reproduce on consecutive runs), they nonetheless allow to identify the corresponding regions on the phase diagram, with more clustering in ferromagnetic phase, checkerboard in antiferromagnetic, and random patterns in paramagnetic. Bigger clusters are observed in the vicinity of the phase transition, and the system becomes progressively more random for higher temperatures.

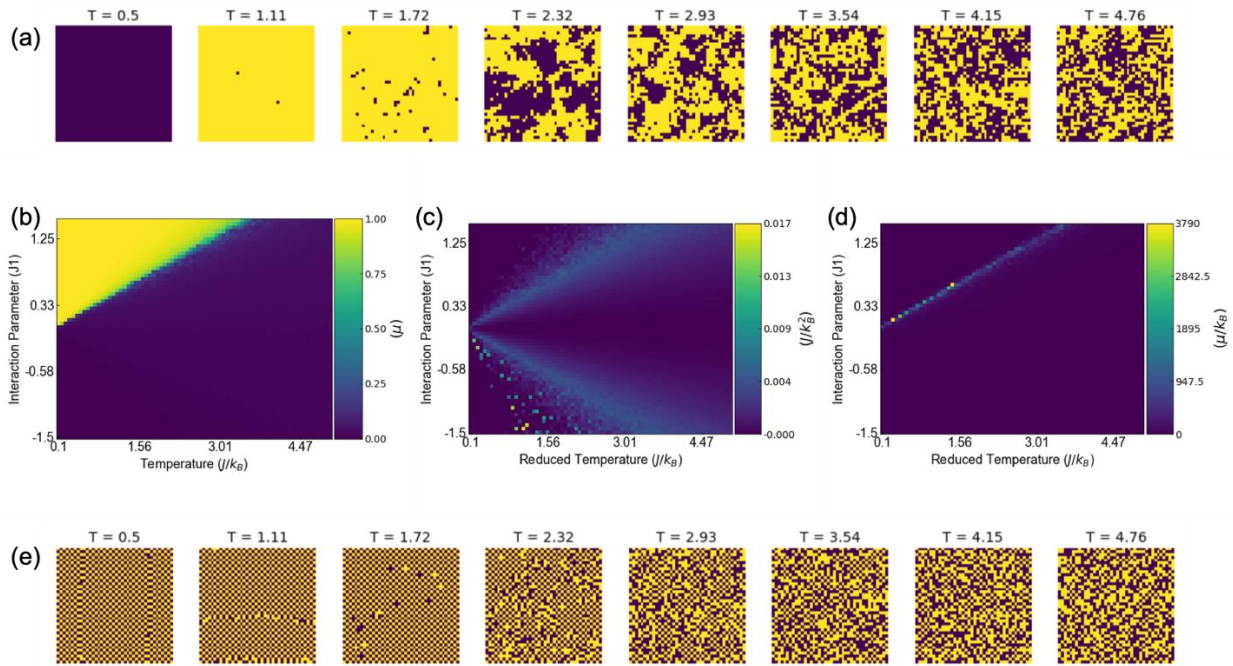

Figure 2. (b) Net absolute magnetization per site, (c) Specific heat, (d) Susceptibility as a function of temperature for the whole parameter space of interaction parameter (J1). (a) and (e) configurations corresponding to ferromagnetic and anti-ferromagnetic regions at different reduced temperatures

These considerations suggest that the experimental images of a material at a single given temperature can be used to extract the interaction parameters from the spatial distributions of the observed microscopic states, and this information can be used to generalize the behavior over a finite temperature interval. Here, we explore the potential for such a reconstruction for the Ising model interaction parameter ($J_1$) extracted from the microscopic observation of spin configurations. We refer to the simulation given configuration as a pseudo-experiment. During the reconstruction, we have access to the spin configurations, but not the Ising model parameter. To determine the unknown model parameter, $J_1$, we use the forward model to generate the spin configurations of the pseudo-experiment case. The value of $J_1$ of the simulations that gives spin configurations with the closest match to the pseudo-experiment is the parameter we seek.

Monte Carlo simulations were carried out for sixty distinct interaction parameters in the range $J_1 \in \{-1.5, 1.5\}$. To achieve the reconstruction, we establish pseudo-experimental results which are 400 distinct microstates for $J_1 = 0.65$ and bond-disorder $= 0.1$ system at a specific reduced temperature. Snapshots from the last 400 steps of the MC simulations are used as experimental observables. The relative frequencies of local (nearest-neighbor) configurations at the specified reduced temperature for all the interaction parameters simulated are compared to the limited pseudo-experiment dataset to estimate the interaction parameter. To compare these two systems (i.e. the model and pseudo-experiment), we employ the statistical distance metric, which is used to quantify the distinguishability of a pair thermodynamic systems, and is formally defined as,[29]

$$s = arccos\left(\sum_{i=1}^{k} \sqrt{p_i}\sqrt{q_i}\right) \qquad (3)$$

Where $p_i$ and $q_i$ are the probabilities of finding local configuration $i$ in the pseudo-experiment and model simulations, respectively, and the summation runs over all $k$ local configurations. Ideally, as the distance goes to zero during the optimization, the measurements performed on the target and model systems become indistinguishable based on statistical hypothesis testing. The metric is unique in that it takes into account the statistical significance of different features in the collected distributions. This allows our procedure to separate even weak signal, such as thermal fluctuations, from the statistical noise naturally present in sampling from a thermodynamic system. Capturing of such fluctuations by the optimized model is then critical for its predictive abilities.

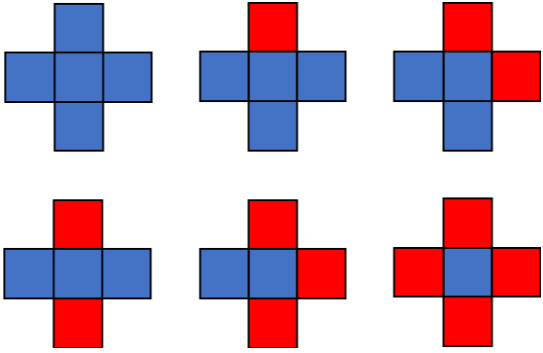

**Figure 3.** An exhaustive set of six local configurations considered for comparing two different macrostates of a system, blue and red indicate opposite spins.

Here we compared the two systems based on the set of six local configurations shown in Figure 3. The optimal parameter $J_1'$ was selected as leading to the shortest statistical distance between the two histograms of these configurations. The interaction parameters predicted here used only a finite number of 40*40 microstate configurations from the pseudo-experiment. This showcases the power of the reverse-Ising model developed here to extrapolate limited experimental data to predict behavior of the system.

**Uncertainty quantification**

Since sampling in any real experiment is limited, the key question of exactly how many samples are required to achieve a certain tolerance on the inferred parameters is an important one. Towards this aim, we utilize two methods to determine the uncertainty in the reconstructed parameter $J_1'$. As a first method, we consider using the statistical distance metric directly: that is, statistical distance as a measure of separation between two multinomial distributions in the probability space is used directly to evaluate the likelihood that a sample is generated from a given model or target distribution. In the limit of large sample, where each sample corresponds to an individual local configuration, the likelihood function attains the shape of the normal distribution centered around the limiting distribution with variance equal to 1/4 as a result of the variance-equalizing transformation built into the probability space with metric $s$.[30, 31] The log-likelihood of a sample at a distance $s$ from the limiting probability distribution is then proportional to $-2ns^2$, where $n$ is the number of samples (individual local configurations). As the distance $s$ increases the likelihood decreases and it does so with a linear dependence on the number of samples $n$. Each value of $J$ is associated with a statistical distance $s$, and therefore from the above relation we can compute the probability estimate for that chosen value of J. Note that in this case we are only computing point estimates, not the whole distribution. This can be repeated for all J values and then the mean and standard deviation can be computed. We applied this method to a situation where $J_1$ was set to 0.65 with a bond disorder of 10% (microstates as a function of temperature are plotted in Figure **4**(a); the results are shown in Figure 4(b)) and indicate that uncertainty in the parameter continues to increase as the temperature increases, as would be expected.

As an alternative approach we consider an averaging method utilizing the samples at hand. In this case, we divided the parent data set of 400 images into 20 sets of uncorrelated images. Interaction parameters are reconstructed for each subset at a given reduced temperature, and mean, standard deviation and the probability density function of the reconstructed values are reported. The results are shown in Figure 4(c) and indicate that the mean remains consistent and the standard deviation increases with increasing temperature, as to be expected. The same data plotted as a probability density is shown for the two methods in Figure **4**(d) and Figure **4**(e).

The effects of reduced temperature on the applicability of the reverse Ising model are reflected in this analysis. Below the transition temperature, the spin configurations of the system are all perfectly aligned hence it is difficult to estimate the interaction parameters if the pseudo-experimental data is considered at $T<T_t$. The model shows the most optimum predictive capabilities occur at the transition temperature. At temperatures greater than the transition temperature, the spin configurations are more random and lack long-range ordering. It is observed that the reverse Ising model performs poorly at temperatures below the transition temperatures and at temperatures much greater than the transition temperature. This is an important result as it shows that even the random spin-configurations at $T>T_t$ possess some information about the interaction parameters of the system.

Here it is also clear that the likelihood method produces higher uncertainty estimates (by a factor of 2 to 3) than the experiment averaging method. This might be because in the first instance the statistical distance is taken into consideration and reflected in the calculated probability; in the second case, although the statistical distance is utilized for optimization of each J there are no point estimates to the variance of each J value and therefore, when the averages are computed the standard deviation is significantly lower.

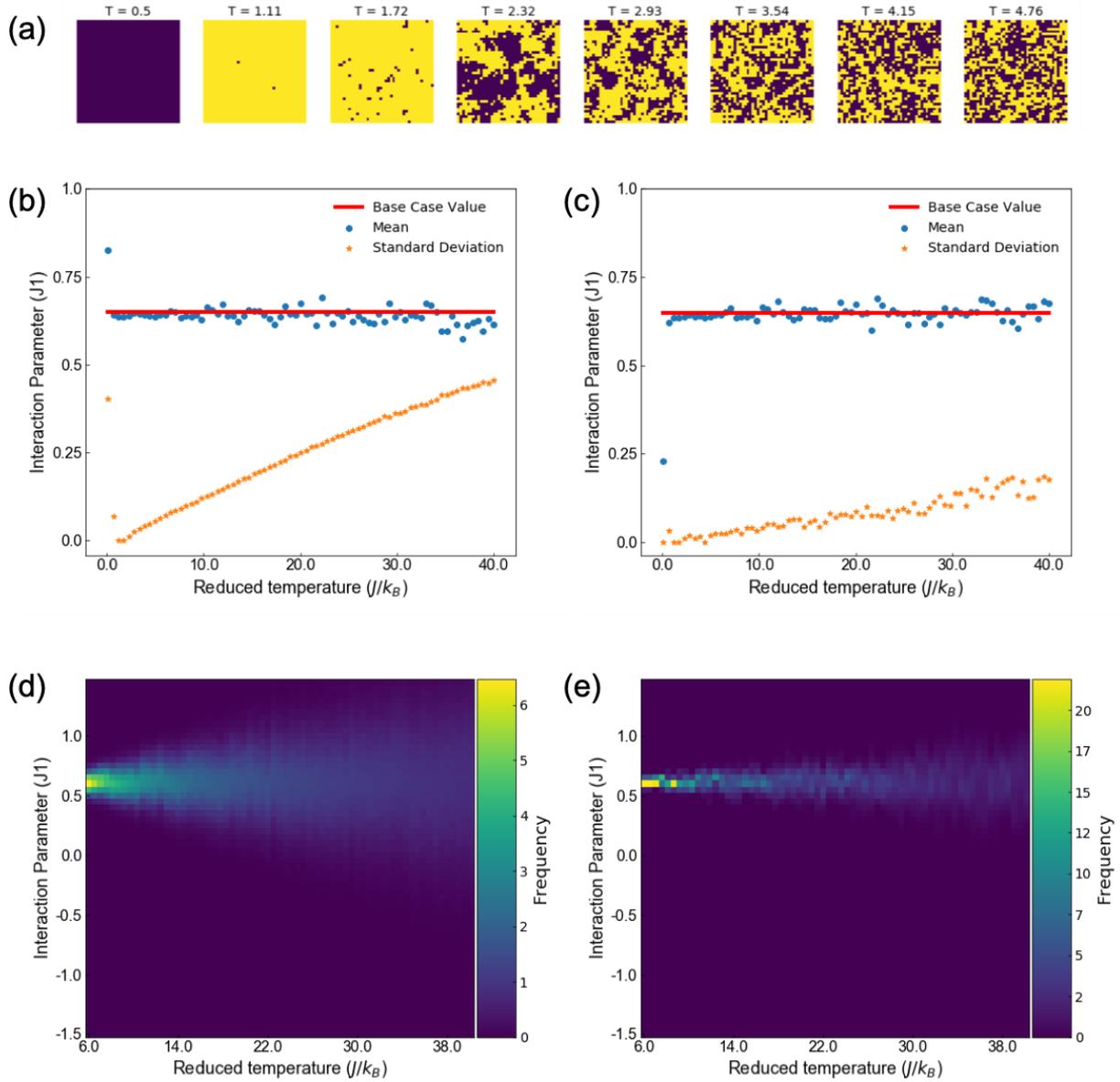

**Figure 4.** (a) Base case microstate configurations at various temperatures ($J_1 = 0.65$, bond disorder = 10%) for the ferromagnetic case. (b) Mean and standard deviation of the likelihood probability of base case at various temperatures. (c) Mean and standard deviation derived from the experimental averaging method. (d), (e) pdf of uncertainty as a function of reduced temperature for the likelihood probability method (d) and the experimental averaging approach (e).

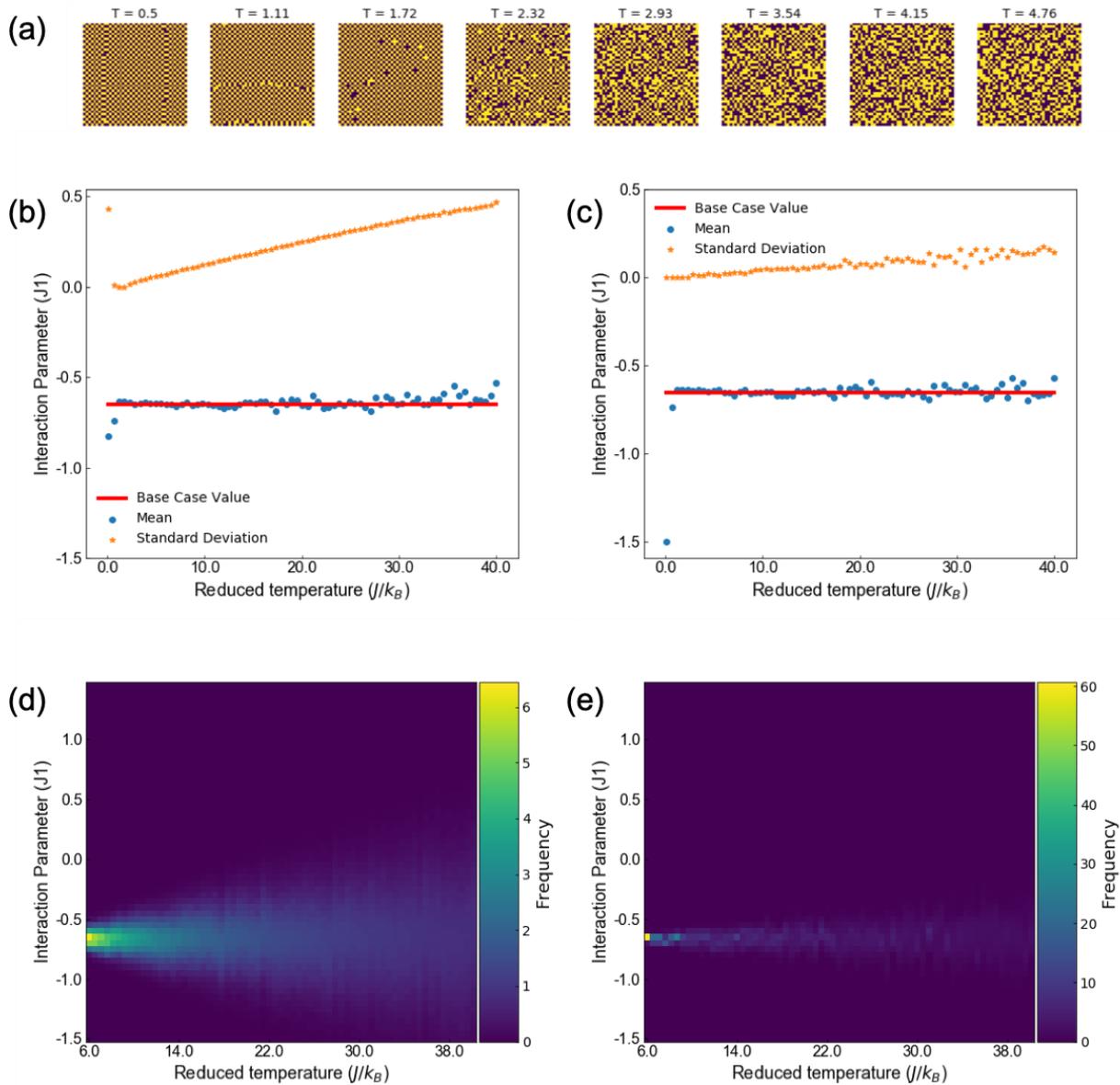

**Figure 5.** (a) Base case microstate configurations for anti-ferromagnetic case ($J_1 = -0.65$, bond disorder = 10%) at various temperatures. (b) Mean and standard deviation of the likelihood probability of base case at various temperatures. (c) Mean and standard deviation derived from the experimental averaging method. (d), (e) pdf of uncertainty as a function of reduced temperature for the likelihood probability method (d) and the experimental averaging approach (e).

A similar analysis is performed in the anti-ferromagnetic region (Figure 5a-e). Temperature dependence of interaction parameter reconstruction for the antiferromagnetic to paramagnetic transition follow similar trends as observed in the ferromagnetic to paramagnetic transition. We conclude therefore that it is possible to reconstruct $J_1$ regardless of sign, at temperatures well above the transition temperature for the 2D Ising model on a square lattice and do so with quantified uncertainty.

**Conclusions:**

Overall, we have theoretically explored the inverse statistical distance-based reconstruction of the generative physical model from the observed microscopic states and perform the associated uncertainty quantification based on log-likelihood and experimental averaging approaches. This analysis suggests that the generative physical model parameters can be extracted well above the transition temperature, with the uncertainty determined by the temperature. In terms of implications for the experiment, this suggests that statistical analysis of the traditionally "bad" experimental objects such as disordered materials can offer insight in underpinning physics, and the fundamental physical mechanisms can be determined well above associated transition temperatures.

Acknowledgements: The work was supported by the U.S. Department of Energy, Office of Science, Materials Sciences and Engineering Division. Research was conducted at the Center for Nanophase Materials Sciences, which is a US DOE Office of Science User Facility.